\documentclass[authoryear,5p]{elsarticle}

\usepackage{lineno,hyperref}
\usepackage{natbib}
\usepackage{multirow}
\usepackage{xcolor}
\usepackage{csquotes}

\journal{Astronomy and Computing}
\begin{document}
\begin{frontmatter}

\title{Multivariate analysis of cosmic void characteristics}

\author[CPPM]{M.C Cousinou\corref{mycorrespondingauthor}}
\cortext[mycorrespondingauthor]{cousinou@cppm.in2p3.fr, CPPM, 163 avenue de Luminy, F-13288 Marseille, France}
\author[Princeton,CPPM]{A. Pisani}
\author[CPPM]{A. Tilquin}
\author[Munchen]{N. Hamaus}
\author[CPPM]{A.J Hawken}
\author[CPPM]{S. Escoffier}

\address[CPPM]{Aix Marseille Univ, CNRS/IN2P3, CPPM, Marseille, France}

\address[Princeton]{Department of Astrophysical Sciences, Princeton University, Peyton Hall, Princeton NJ 08544, USA}

\address[Munchen]{Universit\"ats-Sternwarte M\"unchen, Fakult\"at f\"ur Physik, Ludwig-Maximilians Universit\"at, Scheinerstr. 1, D-81679 M\"unchen, Germany}

\begin{abstract}
The aim of this study is to distinguish genuine cosmic voids, found in a galaxy catalog by the void finder ZOBOV-VIDE, from under-dense regions in a Poisson distribution of objects. For this purpose, we perform two multivariate analyses using the following physical void characteristics: volume, redshift, density contrast, minimum density, contrast significance and number of member galaxies of the void. The multivariate analyses are trained on a catalog of voids obtained from a random Poisson distribution of points, used as background, and a catalog of voids identified in a mock galaxy catalog, used as signal. The classifications are then applied to voids extracted from the Data Release 12 sample of Luminous Red Galaxies in the redshift range 0.45 $\leq$ z   $\leq$ 0.7 from the Sloan Digital Sky Survey Baryon Oscillation Spectroscopic Survey (SDSS BOSS DR12 CMASS). Our results show that the resulting void catalog is nearly free of contamination by Poisson noise. We also study the effect of tracer sparsity and  bias on the classification efficiencies.
\end{abstract} 


\end{frontmatter}


\section{Introduction} \label{S1}
Cosmic Voids are now frequently used as a probe to measure cosmological parameters. They are often defined as large underdense regions (with sizes ranging from tens to hundreds of Mpc*h$^{-1}$, and central densities between 10$\%$ and 50$\%$ of the mean cosmic density) and are found in galaxy surveys as regions of space only sparsely populated by galaxies.
Based on this definition, a large variety of void finding algorithms have been developed. However, most void analyses utilise one of two main classes of void finder. The first class attempts to look for spherical under-densities in the large scale structure~\citep{padilla2005spatial,kitaura2016signatures,sanchez2016cosmic,hawken2017vimos}.
The second class is based on Voronoi Tessellation and the watershed transform and is exemplified by ZOBOV~\citep{neyrinck2008} or VIDE~\citep[][itself based on ZOBOV]{sutter2015vide}. 
An additional third class of void finders~\citep[e.g.][]{hahn2007properties,lavaux2010precision,elyiv2015cosmic} rely on the dynamical properties of galaxy distribution.  

In this paper our study will focus on the characteristics of voids found with ZOBOV-VIDE. Many analyses have been performed on cosmic voids found with this void finder. For example: Alcock-Paczy\'{n}ski tests~\citep{lavaux2012cosmic,sutter2012first,sutter2014measurement,Mao2016faj} or studies of redshift-space distortions (RSD)~\citep{hamaus2015probing,hamaus2016constraints,hamaus2017multipole}. However, as explained in~\citet{neyrinck2008} and \citet{nadathur2015natureI,nadathur2015natureII} the presence of shot noise can lead the ZOBOV algorithm to output some spurious shallow voids. In these analyses different cuts have been used to trim the void catalog in an attempt to prune out voids generated by Poisson noise.  In~\citet{Mao2016faj} the density contrast (see the definition in section~\ref{S2}) was chosen
as a measure of the void significance. 
A central density cut that excludes voids
for which the density around the void center is above 0.2 $\bar{\rho}$, where  $\bar{\rho}$ is the mean density, can also be applied. This cut was initially
used in~\citet{sutter2012public} but such a criterion imposes additional constraints on the central shape of the density profile~\citep{sutter2014sparse}, which is not the subject of this paper. Another density criterion has been suggested in~\citet{nadathur2015natureI}. The void center (or circumcenter) is chosen as the point of intersection of the four lowest density mutually adjacent Voronoi cells in the void. The void is then classified as `spurious' or `genuine' according to whether it is overdense or underdense compared with the mean density, respectively, at the circumcenter. However, voids identified in a Poisson distribution of particles are also underdense. Furthermore, some genuine voids can exhibit overdense cores, which can be used as an argument against this type classification scheme.
In other analyses, the choice has been made to
exclude voids with radii below twice the mean particle separation, this mean particle separation being calculated as a function of redshift~\citep{hamaus2016constraints,hamaus2017multipole}.
However, using the void catalogs described  in section~\ref{S4},
a comparison between the abundance of voids as a function of radius  from SDSS BOSS DR12 CMASS
data with the abundance of voids found in a random galaxy catalog,  shows that keeping only the voids with a radius greater than twice the mean particle separation strongly reduces the size of the sample  (65\% of voids kept)  without removing all of the voids of the random sample (44\% of random voids are not discarded). 

So it seems that these cuts, used individually, do not fully succeed in significantly reducing the contamination due to Poisson noise. The aim of this paper is to find out if a multivariate study, in which the final selection is determined by the differences between the combination of a number of void characteristics, would be more powerful. It is organized as follows: after the introduction, we recall in section~\ref{S2} the main features of the void finder, VIDE, and the meaning of the variables we will use in the multivariate analysis; the multivariate analysis methods are described in section~\ref{S3}; their application to the SDSS BOSS DR12 CMASS galaxy catalog and the effect of the use of this selection on a redshift-space distortion (RSD) analysis are described in section~\ref{S4}, together with the results obtained using the multivariate analysis methods on lower density tracer samples or on a sample of galaxies with a different bias; the final section contains a short discussion of the results together with the conclusion.

\section{Description of VIDE and its main variables} \label{S2}
To identify voids in a tracer catalog we use VIDE~\citep{sutter2015vide,lavaux2012precision} which is an enhanced version of the void finder ZOBOV~\citep{neyrinck2008}.
ZOBOV starts by using a Voronoi tessellation to construct the density field of a discrete distribution of tracers. It divides space into cells around each tracer, with the cell around tracer $i$ defined as the region of space closer to $i$ than to any other tracer. A number density $\rho_{i} = 1/V_{i}$, where $V_{i}$ is the volume of the Voronoi cell around the tracer, is assigned to this tracer. The second step partitions the set of cells into small voids around each density minimum.
The watershed transform is used to join zones together to form the final voids~\citep{platen2007cosmic}. It merges zones starting from the minima of the density field. Adjacent zones are added  if the minimum density along the ridge separating them from the void is smaller than $ 20 \% $ of the mean particle density~\citep{neyrinck2008,sutter2015vide}. 

VIDE outputs variables for each void that describe their physical characteristics, such as size and depth,
and gives the list of tracers by which the void is defined. 
\begin{itemize}

\item Three variables give the size: a volume, $V$, defined as the sum of the volumes of the Voronoi cells that contribute to the void; an effective radius, $ r_v = ( \frac{3}{4\pi}V)^{1/3}$; and a normalized volume ($\hat{V}$) which is $V$ divided by the volume occupied by a mean-density particle. As these three variables are fully correlated, we used only one: the normalized volume. 

\item The depth of the void is described by the core density variable ($ \rho_c $), the density of the largest Voronoi cell in the void. 

\item The density contrast ($r$), the ratio of the minimum density of the ridge separating the void from adjacent zones to the core density of the void.

\item As pointed out in~\citet{neyrinck2008}, a ZOBOV void is simply a density minimum with a depression around it. Therefore, when applied to sparse or noisy data, ZOBOV returns many shallow voids. In~\citet{neyrinck2008} the cumulative probability function $P(r)$ of the density contrast $r$ was fitted for two cubic Poisson simulations as:  $P(r) = e^{-5.12(r-1) -0.8(r-1)^{2.8}} $. This $P(r)$ states the probability for a given void to arise from a Poisson distribution, when only its density contrast is taken into account. We name this variable the `contrast significance'. This significance decreases as the density contrast increases.

\item  The number of tracers that define the void ($N_t$). 

\item Finally, as the size of the void is slightly linked to its redshift ($z_v$) (\citet{sheth2004hierarchy} or see values in Fig.11 of~\citet{hamaus2017multipole}),  we include it in the set of variables used for the multivariate analysis.

\end{itemize}

\section{Classification of voids with a Multivariate Data Analysis} \label{S3}
The two methods described here are multivariate classifications based on supervised training.
Multivariate classification  is a discriminant analysis to separate events into classes, based on differences between the distribution of variables, given as input. There are two steps to the analysis. In the first supervised learning step, the classification method must be trained on known samples which already provide the outcome: a sample made from a background process  and a sample made from a signal process. Each of these samples is split into a training sample and a test sample. During this step the classifier computes optimal values of the weights in order to maximize signal to background separation. The result of the training is evaluated on the test sample. The second step is the application of the output of the training to the sample we want to classify: the weights calculated during the training step are used to classify objects in the unknown sample.
A MultiVariate data Analysis (MVA) makes use of hidden correlations between the variables to combine several discriminating variables into one final discriminator. This gives better separation than cuts on individual variables alone. In our analysis we used two well known classification techniques. Namely the Boosted Decision Tree, and an Artificial Neural Network technique: the Multi-Layer Perceptron.

\subsection{The Boosted Decision Tree (BDT)} 
\begin{figure*}
	\includegraphics[width=\columnwidth]{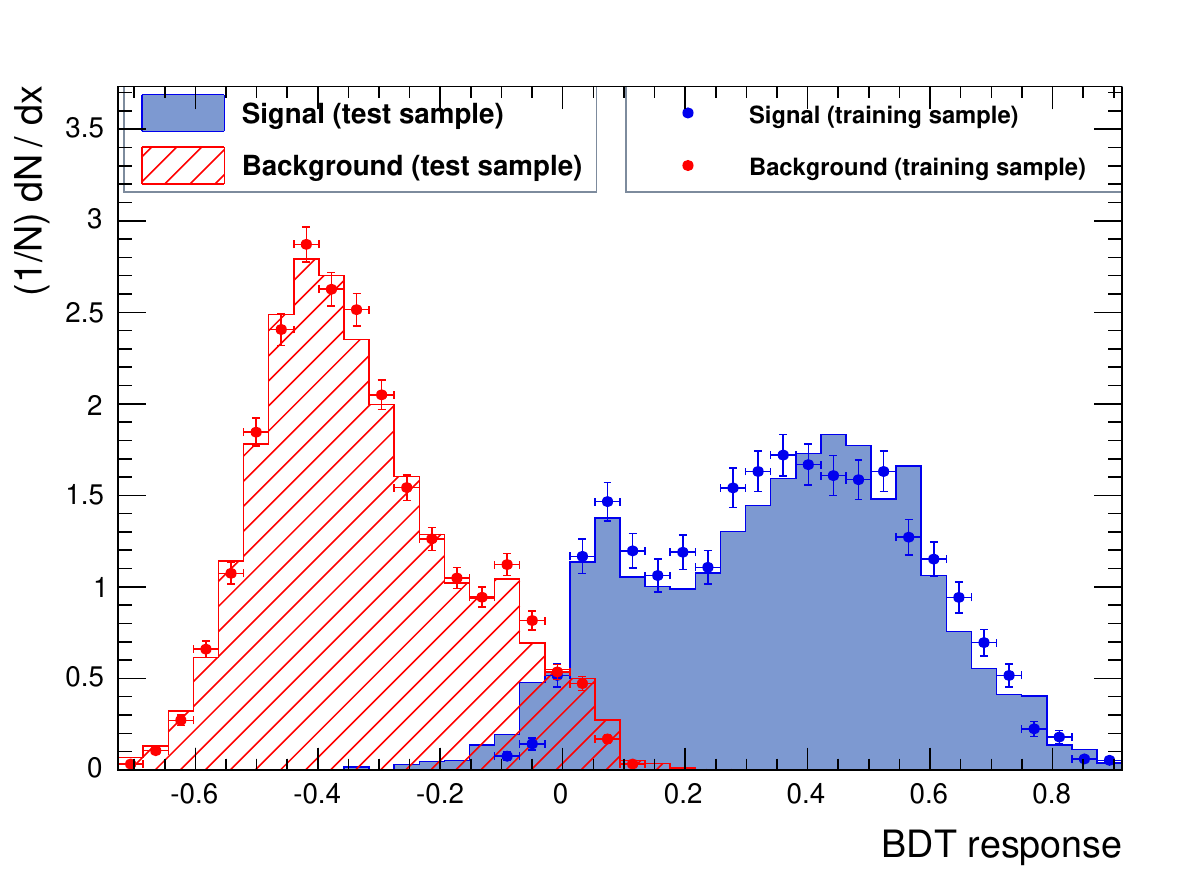}
 	\includegraphics[width=\columnwidth]{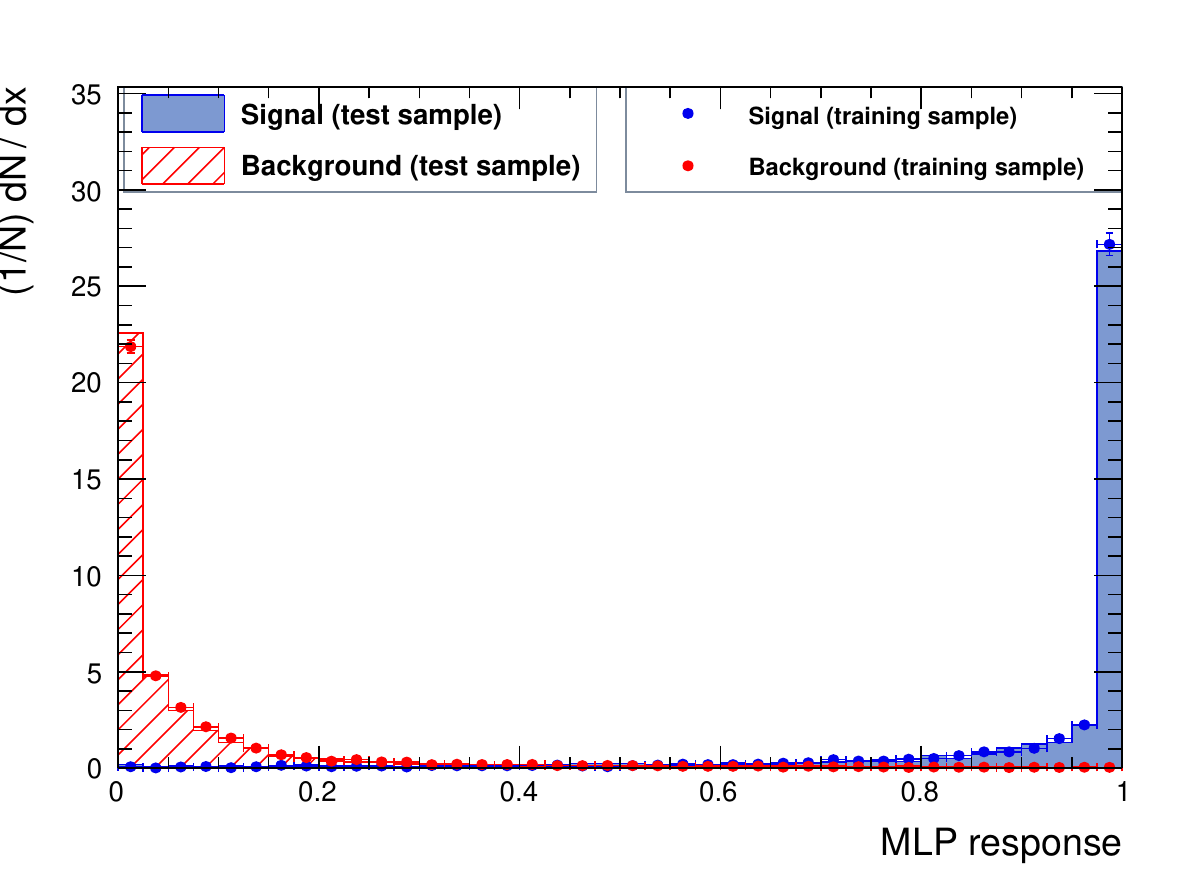} 
    \caption{Responses of the BDT and MLP classifications on the left-hand and right-hand plots respectively. The dots with the error bars correspond to the training sample responses and the histograms to the test sample. The results for the signal are plotted in blue color and those of the background in red color.}
    \label{fig:response} 
\end{figure*}

A Decision Tree consists of a consecutive set of questions (nodes), with each question having only two possible answers. At each level the question depends on previously given answers. The choice of node criterion is made by maximizing separation gain between nodes. The gain can be computed in different ways, for example: $ gain \cong p (1-p)$ where $p$, the purity, is the proportion of signal in a sample containing signal and background. The final node (called the `leaf')  is reached after a given maximum number of nodes or to keep a minimum number of training events. At the end of the training of the tree, each leaf is classified as signal or background with an associated purity value: $P_{S}=\frac{S}{S+B}$ or $P_{B}=\frac{B}{S+B}$, $S$ and $B$ being, respectively,  the number of weighted events from the signal and background samples.
Single trees are not very powerful, so the BDT tool uses the Random Forests method, which combines many different trees. The most commonly used method to train the Random Forests is boosting (AdaBoost; \citet{freund1999short}) which enhances the weight of misclassified events and reduces the weight of correctly classified ones after each training. This ensures that future trees learn better. The trees are finally combined into a single classifier which is given by a weighted average of the individual decision trees. For event $i$, the output of the BDT is $T(i) = (1 /  \sum\limits_{K=1}^{N_{trees}} {\alpha_{K}} )  \sum\limits_{K=1}^{N_{trees}}{\alpha_{K} T_{K}(i)} $ where $\alpha_{K} = \ln{\frac{1-\epsilon_{K}}{\epsilon_{K}}}$  and $\epsilon_{K}$ is the proportion of misclassified events after the training of tree $T_{K}$ and  $T_{K}(i)$ the result of tree $K$. The output of this BDT tool is a number between -1 and 1. As can be seen in Fig.~\ref{fig:response}, the output values corresponding to the background and the signal are mainly clustered in the intervals [-1,0] and [0,1], respectively. We shall henceforth refer to this number as the "BDT response".

\subsection{The Multi-Layer Perceptron (MLP)}
An Artificial Neural Network is a collection of interconnected units (nodes) called artificial neurons. There are $N$ input variables $(x_i)_{i=1,N} $ with associated  weights which will constitute the input-layer of N neurons. The input of each neuron consists, most of the time, of  the weighted sum $y$ of the input variables. The output is $f(y)$, where $f$ is a non linear function called the activation function. The most frequent non linear activation functions are the hyperbolic tangent and sigmoid functions. The complexity of a neural network with $n$ neurons, which could have $n^{2}$ connections, is reduced by the organization of these neurons into layers. Only direct connections from one given layer to the next are allowed. This kind of artificial neural network architecture is known as a Multilayer Perceptron. The first layer is the input-layer and the final one the output-layer.  Other layers (often only one layer) are referred to as hidden-layers. The weights $(w_{ij})$ where  $iÊ\in [1,N]$ and $j \in [1,n_{h}]$, $n_{h}$ being the number of hidden-layers, are computed during the training phase in such a way as to minimize the difference between the output and the desired value. The output of this MLP tool is a number taking values between 0 and 1 :
$  \sum\limits_{j=1}^{n_{h}}{f(\sum\limits_{i=1}^{N} x_{i} w_{ij}^{(1)}})  w_{j1}^{(2)}$  where $w_{ij}^{(1)}$ is the weight between input-layer neuron $i$ and hidden-layer neuron $j$ and $w_{ij}^{(2)}$ is the weight between the hidden-layer neuron $j$ and the output neuron.
 As can be seen in Fig.~\ref{fig:response}, the output values corresponding to the background and the signal are mainly clustered in the intervals [0,0.5] and [0.5,1], respectively. We shall henceforth refer to this number as the "MLP response".

\subsection{Application of the classification}
When using MVA classifiers, special care must be taken to avoid  overfitting, especially in the BDT case.  Overfitting happens when the
classifier overfits data, so it ends up looking at features peculiar to the training sample. It can occur when a machine learning problem has too few degrees of freedom. For example, if too many parameters of an algorithm are adjusted to too few data points. Overfitting leads to an apparent increase in the classification performance on the training sample and to a decrease in effective performance when measured on an independent test sample. Therefore the signal and background samples are split into a training and a test sample of equal size. When there is comparable performance on the training and test samples this means that there is no overfitting. This can be checked in Fig.~\ref{fig:response}, which shows a comparison of the BDT or MLP responses from the signal and the background sample. Additionally, the toolkit we used for this analysis provides performance estimators which should be similar for the training sample and for the test sample when there is no overfitting. We find that this is the case in our analysis.

For this study, we used the Toolkit for Multivariate Data Analysis (TMVA)~\citep{Hocker:2007ht} from the ROOT~\citep{brun1997root} package. ROOT provides a machine learning environment for the processing of data using multivariate classification techniques. We used the recommended parameters for these MVA techniques~\citep{Hocker:2007ht}. For the BDT method, the number of trees in the forest has been adjusted to 700 to optimize classification and to avoid overfitting. For the MLP method we used, as suggested, one hidden layer with a number of neurons equal to the number of input variables + 5 (so 11 in our case), 600 training cycles and the hyperbolic tangent as activation function. We also considered a more complex structure (2 hidden layers and 24 neurons) and the use of a sigmoid function instead of the hyperbolic tangent. We found that the differences in the output of the neural network were negligible.

\section{Multivariate analysis of the SDSS BOSS DR12 CMASS void sample} \label{S4}
\subsection{Description}\label{des}
As mentioned in the introduction, the VIDE algorithm can output a number of shallow voids due to shot noise. The multivariate data analysis, described here, was motivated by the need to know how many voids of this kind were in a void catalog used for cosmological analysis~\citep{hamaus2017multipole} and by the wish to enhance signals with low signal to noise for future analysis. The data void catalog was extracted from the publicly available Data Release 12~\citep{reid2015sdss,alam2017clustering} catalog of Luminous Red Galaxies from the SDSS-III Baryon
Oscillation Spectroscopic Survey~\citep{dawson2012baryon} (BOSS), in the redshift range 0.45 $ \leq$  z  $ \leq$  0.7.
To train and test the MVA we used a catalog of voids found in a mock galaxy catalog as signal and a sample of voids extracted from randomly distributed points with the same mean number density as in the mock as noise (which we shall refer to as Poisson noise voids). We used the official BOSS DR12 mock and randoms, to replicate the data as closely as possible. The mock has been built from MultiDark PATCHY BOSS DR12 light-cones (with the cosmology: $\Omega_{m}$=0.307, $\Omega_{\Lambda}$=0.693, $\Omega_{b}$=0.048, $\sigma_{8}$=0.829 and h=0.678) and the randoms were generated to have the same angular and redshift selection function as the target galaxies of the survey~\citep{kitaura2016clustering,rodriguez2016clustering}. The signal and background samples contain $5,820$ and $13,607$ voids respectively. The background sample is free of real cosmic voids. The signal sample, however, being the output of the VIDE algorithm, could contain a mixture of genuine and spurious voids. Since the background sample contains only spurious voids, the fact that the signal sample could contain some background voids makes the classification less efficient than if we were able to obtain a pure signal sample.

The distributions of the six input variables, for the signal and background samples, are shown in Fig.~\ref{fig:inputvar}. In this figure we can see that the distributions for the variables from the signal and the background samples show important differences, in particular between the core density $\rho_c$ and contrast significance $P(r)$. These differences imply that, on average, the voids associated with the Poisson noise are smaller, shallower (their contrast significance $P(r)$ is higher) and have a higher minimum density. They are also defined by a smaller number of particles. 
The output responses of the two MVA methods for the training and test samples, as shown in Fig.~\ref{fig:response}, are in good agreement and show no evidence of overfitting.
Curves showing the signal and background efficiencies as a function of the BDT response and the MLP response are displayed in Fig.~\ref{fig:efficiency} (left-hand and right-hand plots respectively) together with the significance $\frac{S}{\sqrt{S+B}}$, where $S$ and $B$  are the number of voids in the signal and background samples which pass the classifier cut, respectively. The background efficiency  is defined as the fraction of voids in the background sample which survive a given classifier cut. As the background sample is a pure sample of Poisson noise voids, this is the efficiency for spurious voids to pass the cut. In a similar way, we refer to the proportion  of voids in the sample taken as signal (voids found in the mock galaxy catalog) which pass the classifier cut, as the signal efficiency. As mentioned earlier, this sample contains a mixture of cosmic voids and a small fraction of Poisson noise voids.  So this efficiency should not be taken to be the cosmic void efficiency. However, when we apply a cut on a classifier to purify the sample of voids from the mock, we will reduce the proportion of Poisson noise voids. Therefore the cosmic void efficiency is greater than the signal efficiency. The cut which gives the best signal efficiency with the highest background rejection is given by the highest value of the significance. The results are summarized in Tab.~\ref{tab:effandbck}. The results obtained with the two methods are very similar: the signal efficiency is close to 96$\%$ with a background contamination of the order of 5$\%$.

The BOSS DR12 data void catalog on which we apply these MVA classifications contains $4,455$ voids. The distributions of the six input variables of this data sample are compared to those of the mock sample in Fig.~\ref{fig:compinputvar}. One can see that the characteristics of the voids from the mock show a high resemblance to, but do not exactly reproduce, those of the voids found in the data catalog. This is as expected. We can see, by comparing Fig.~\ref{fig:inputvar} with Fig.~\ref{fig:compinputvar}, that the differences between the characteristics of Poisson noise voids and data voids are slightly less pronounced than those between the Poisson noise voids and the mock voids. For example, the values of the normalized volume $\hat{V}$ are shifted to smaller values and the values of the contrast significance $P(r)$ and of the core density $ \rho_c $ to higher values. The differences between the distributions of the random void and data void characteristics are therefore decreased. The responses of the BDT classification (left plot) and the MLP classification (right plot) are displayed in Fig.~\ref{fig:compdatatest}, for the data void sample and the mock test set.  There is generally a good agreement between the responses in the mock test set and the data with respect to the background. Nonetheless, there are some minor differences. This is to be expected since the MVAs were trained on the mock. The cosmic web in the mock may be slightly different from that in the data. The subtle difference in the shape of the mock and data response distributions would imply that the separation of Poisson voids from cosmic voids is slightly more efficient in the mock than in the data. The cuts on the MVA responses, given in Tab.~\ref{tab:effandbck}, applied to data  give an efficiency equal to 93 $\%$ in both cases (compared with 95.7$\%$ for the mock).

\subsection{Effect on an RSD analysis}\label{rsd}
To see to what extent the contamination of Poisson voids can affect the cosmological information extracted from voids, 
we have performed a measurement of the RSD  parameter $\beta$ on our SDSS BOSS DR12 CMASS void catalog~\citep{hamaus2017multipole},  by computing the multipoles of the void-galaxy cross-correlation function (more details in appendix). The parameter $\beta$ is obtained from a fit of the ratio of the quadrupole to the difference between the monopole and the cumulative average monopole. In Fig.~\ref{fig:multipoles}, the quadrupole and monopole of the void-galaxy  cross correlation function, as a function of the void-galaxy distance (normalized to the void radius $r/r_v$), for the mock voids and the data voids are plotted together with those of the random void catalog. The monopole and quadrupole of the cross-correlation function of the mock voids and the data voids are very similar, as expected. The quadrupole values for the Poisson noise void-galaxy cross-correlation function are nearly equal to 0 for every value of $r/r_v$. As there are emptier spaces present in the random distribution, the void finder identifies those underdense regions as voids, and the monopole distribution shows an attenuated void-profile shape. 
So, if the SDSS BOSS DR12 CMASS void catalog were contaminated by spurious Poisson-type voids, it would modify the shape of the monopole and quadrupole of the cross-correlation function and thus the value obtained for the $\beta$ parameter.

The fact that the SDSS BOSS DR12 CMASS catalog is free from Poisson noise void contamination can be seen in Fig.~\ref{fig:monopoles_afterTMVA} and~\ref{fig:quadrupoles_afterTMVA}. In these figures, the monopole and quadrupole of the void-galaxy cross-correlation function are shown, with and without MVA cuts. The result of the analysis without applying MVA cuts is shown in the left-hand plots. In the right-hand plots, the differences between the multipoles with or without MVA cuts are displayed. We made three cuts: a cut on the BDT response (BDT response $\geq$ 0.  giving 4,060 voids); a cut on the MLP response (MLP response $\geq$ 0.45  giving 4,142 voids); and a tighter cut on the MLP response (MLP response $\geq$ 0.8  giving 3,603 voids). In the right-hand plots, the one sigma error is shown as a shaded area. The changes due to the different MVA cuts are all within the one-sigma uncertainty of the data.

\subsection{Dependence on tracer sparsity and bias} \label{sp}

  To investigate the effect of shot noise in the galaxy sample, we repeated this MVA study on two lower density tracer samples, corresponding to about 1/5 ($200,000$ galaxies) and 1/10 ($100,000$ galaxies) of the density. For this purpose we randomly subsampled the mock file and the random file. As the number of voids obtained in these subsamples is small (about $1,200$ and 600 for the two densities respectively), we repeated this sub sampling on several mocks and randoms sets and concatenated the resulting void catalogs. In both cases the separation between the signal and the background decreased. In Fig.~\ref{fig:100K_responses}, the BDT and MLP responses for the subsample of $100,000$ galaxies are shown. The separation power between the mock voids and the Poisson noise voids weakens with the tracer sample density as can be seen in Tab.~\ref{tab:effandbck2}. As an example, the signal efficiency equal to 95.8$\%$ for the full mock sample decreases to 91.4$\%$ for the lowest density subsample with a background contamination growing from 5.3$\%$ to 40.6$\%$.
  
  We have also performed this MVA study on a subsample of tracers with a different bias. We selected galaxies with $log(M^{*}) \geq$ 11.5 from the mock catalog, where $M^{*}$ is the stellar mass in units of solar mass.  We resample the random catalog in order to get the same redshift distribution as the tracers after the $M^{*}$ cut. We then extracted a subsample of the same size as the mock after the cut. As in the previous case, the void statistics are small so we repeat this procedure for several mocks and random subsamples. The BDT and MLP responses are plotted in Fig.~\ref{fig:MSTAR_responses} and the signal efficiencies and background contaminations are given in Tab.~\ref{tab:effandbck3}.  We observed, also in this case, a decrease in the performance of the classifiers.

\section{Discussion and Conclusion} \label{S5}
We have performed two multivariate analyses, using the ROOT TMVA package. We have trained them on void catalogs obtained from a mock galaxy catalog (signal) and a random catalog (background). We have applied the resulting classifications to the SDSS BOSS DR12 CMASS void sample. We performed a multivariate analysis on void characteristics such as the normalized volume, the core density, the density contrast, the contrast significance of a void, the number of tracers defining a void, and the void redshift. We have shown that the SDSS BOSS DR12 CMASS void sample is nearly free of Poisson noise type voids. Indeed, when applying a cut corresponding to a MLP response $\geq$ 0.45, the efficiency for the cosmic voids is greater than 93$\%$ in BOSS data with a spurious void contamination equal to 4.5$\%$. Similar results are obtained when we apply a cut on the BDT response (BDT response $\geq$ -0.02) instead of a cut on the MLP output.

\begin{figure*}
	\begin{center}
	   \includegraphics[width=0.8\textwidth]{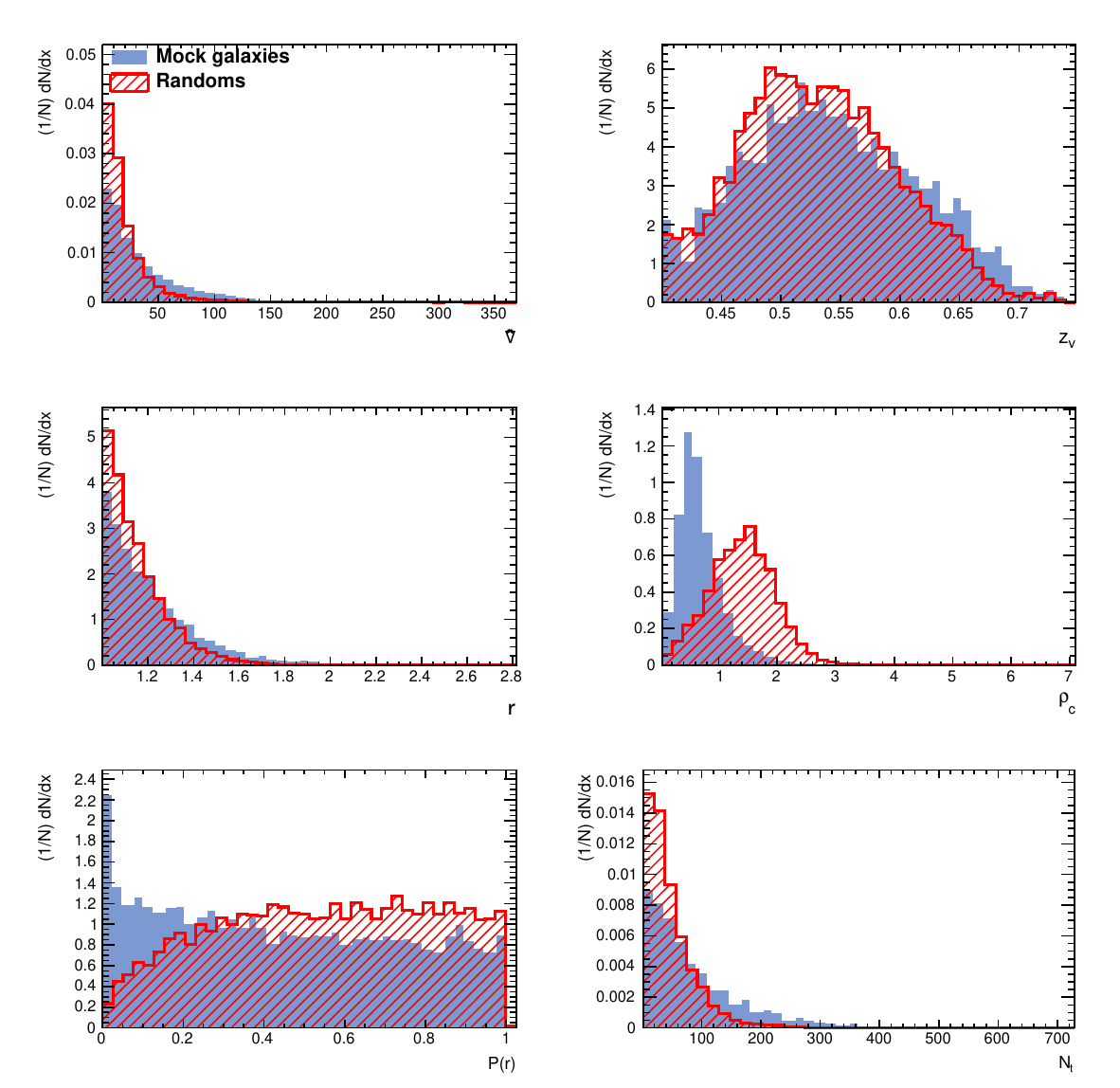}
	\end{center}
    \caption{Distributions of the variables used to discriminate the signal (full blue histograms) from the background (dashed red histograms). From left to right and top to bottom, the variables are the normalized volume ($\hat{V}$), the redshift ($z_v$), the contrast of density  ($r$), the lowest density in the void ($\rho_c$), the contrast significance of a void ($P(r)$) and the number of galaxies defining a void ($N_t$).}
    \label{fig:inputvar}         
\end{figure*}

\begin{figure*}
	 \includegraphics[width=\columnwidth]{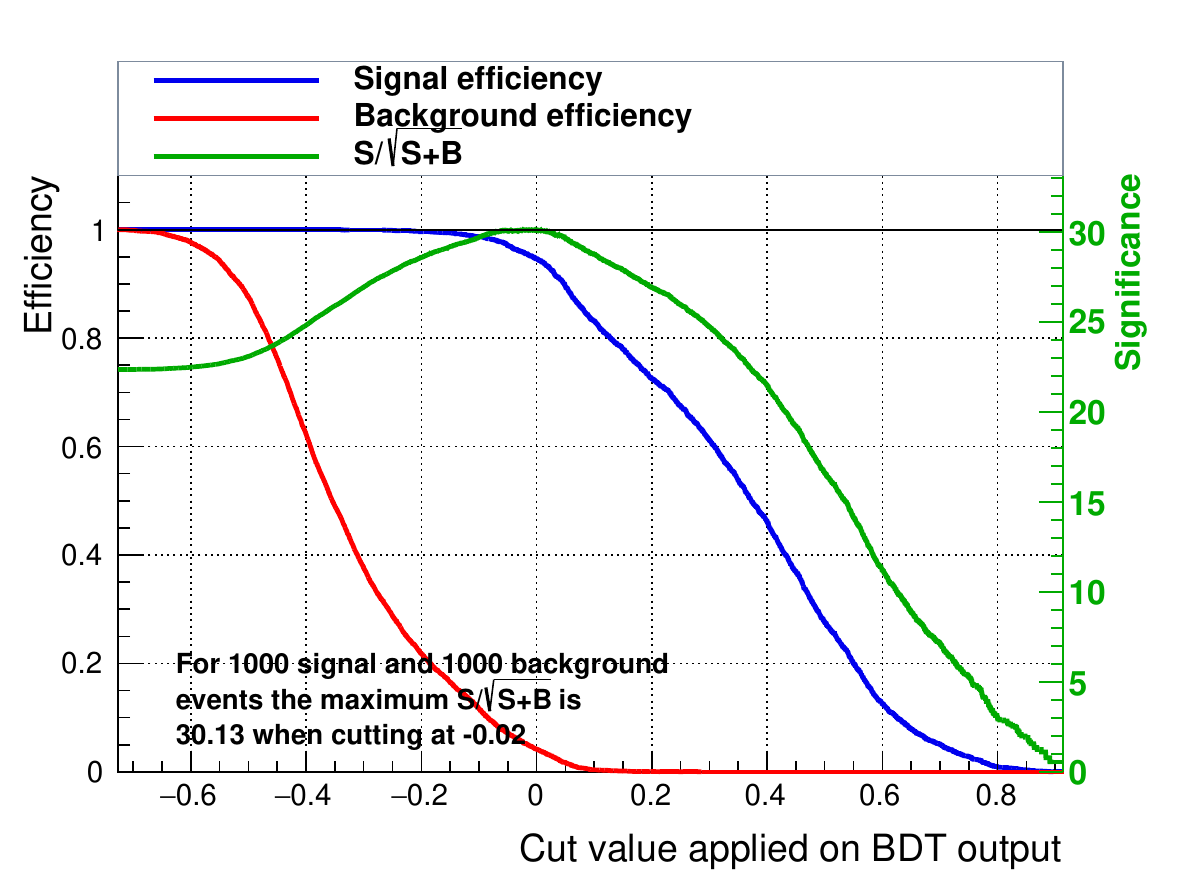}
	 \includegraphics[width=\columnwidth]{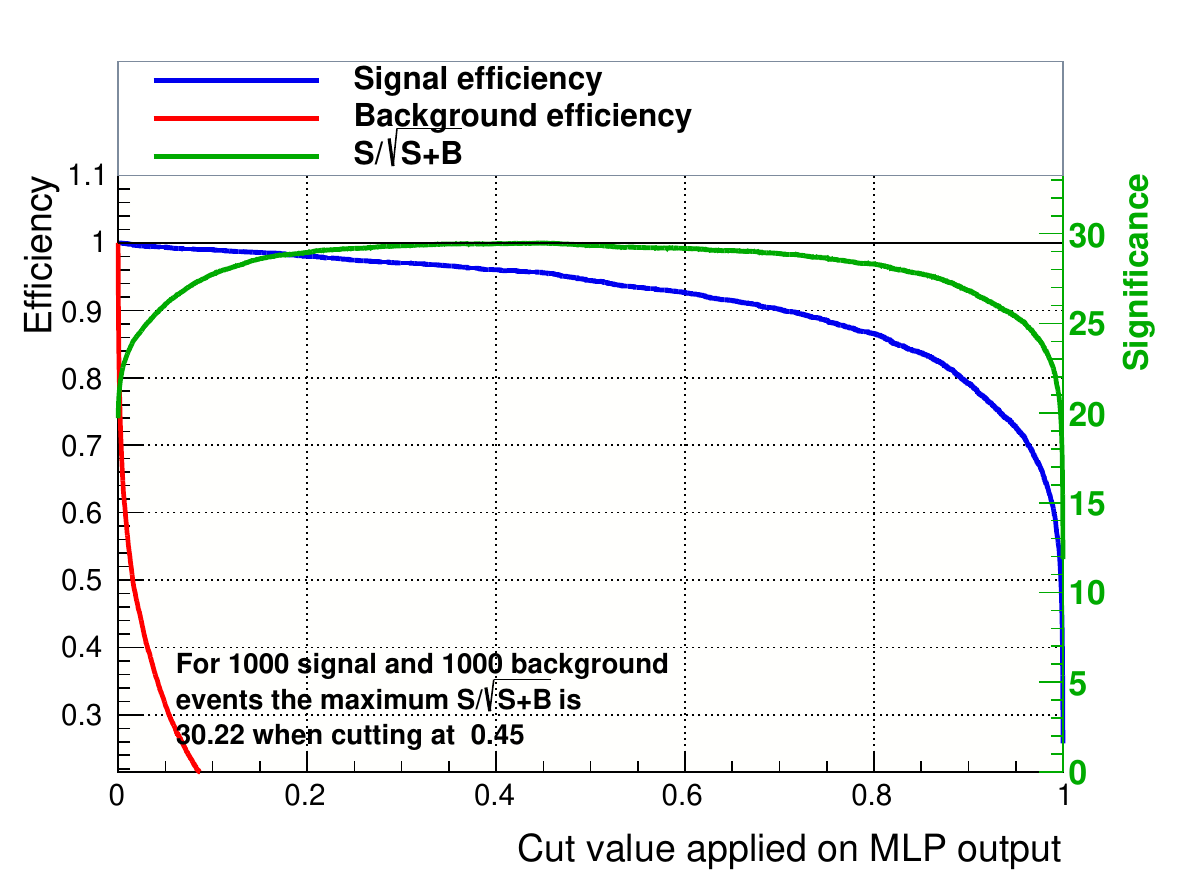}
	     \caption{Efficiencies of the signal (blue curve), of the background (red curve) and significance (green curve) as a function of the BDT response (left-hand plot) and of the MLP response (right-hand plot)}
    \label{fig:efficiency} 
\end{figure*}

\begin{figure*}
	\begin{center}
	   \includegraphics[width=0.8\textwidth]{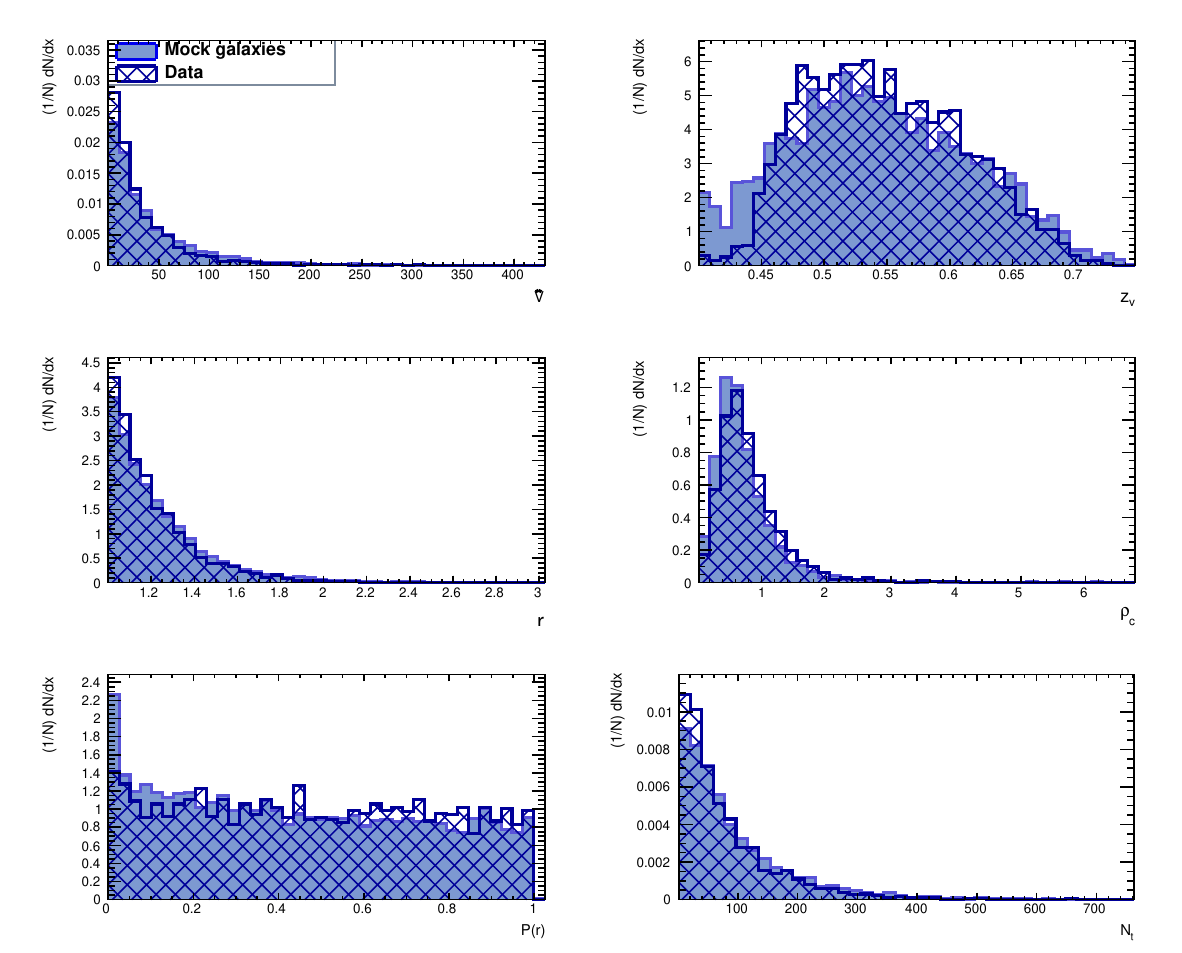}
	\end{center}
    \caption{Comparison of the characteristics of the voids found in the mock catalog (full blue histograms) and the voids from the DR12 CMASS galaxies catalog (hashed blue histograms). From left to right and top to bottom, the variables are the normalized volume ($\hat{V}$), the redshift ($z_v$), the contrast of density ($r$), the lowest density in the void ($\rho_c$), the contrast significance ($P(r)$) and the number of galaxies defining a void ($N_t$).}
    \label{fig:compinputvar}
\end{figure*}

\begin{figure*}
	\includegraphics[width=\columnwidth]{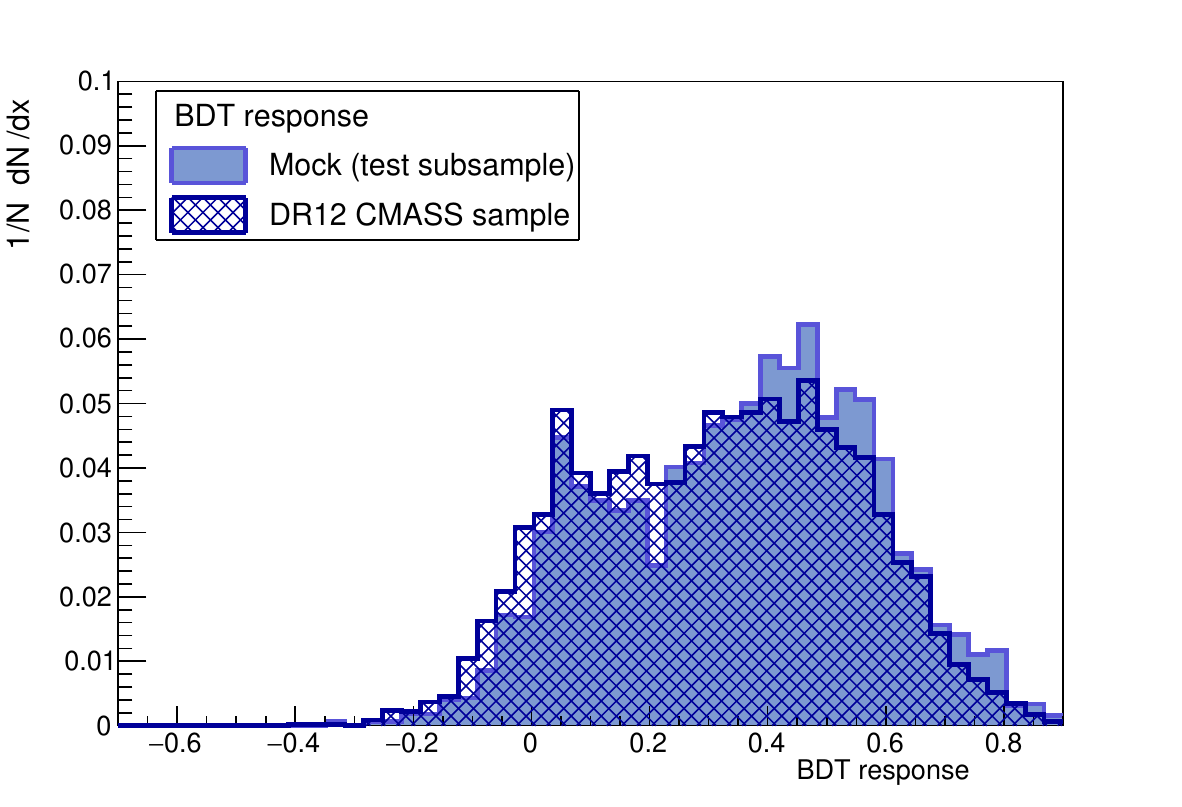}
	\includegraphics[width=\columnwidth]{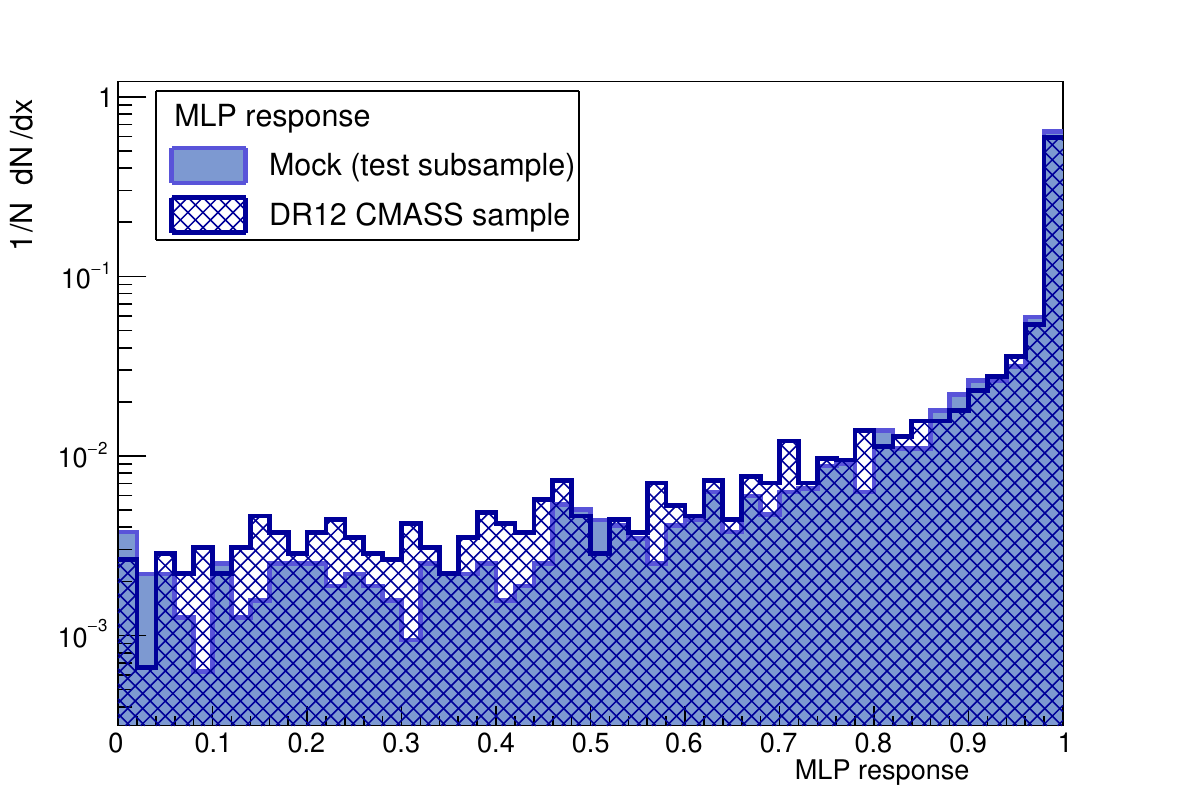}    
    \caption{Comparison of the BDT response (left-hand plot) and the MLP response (right-hand plot) for the voids obtained from the DR12 CMASS data sample (hashed blue histogram) and the voids from the part of the mock sample used as test (full blue histogram).}
    \label{fig:compdatatest}
\end{figure*}

\begin{figure*}[h]
	\includegraphics[width=\columnwidth]{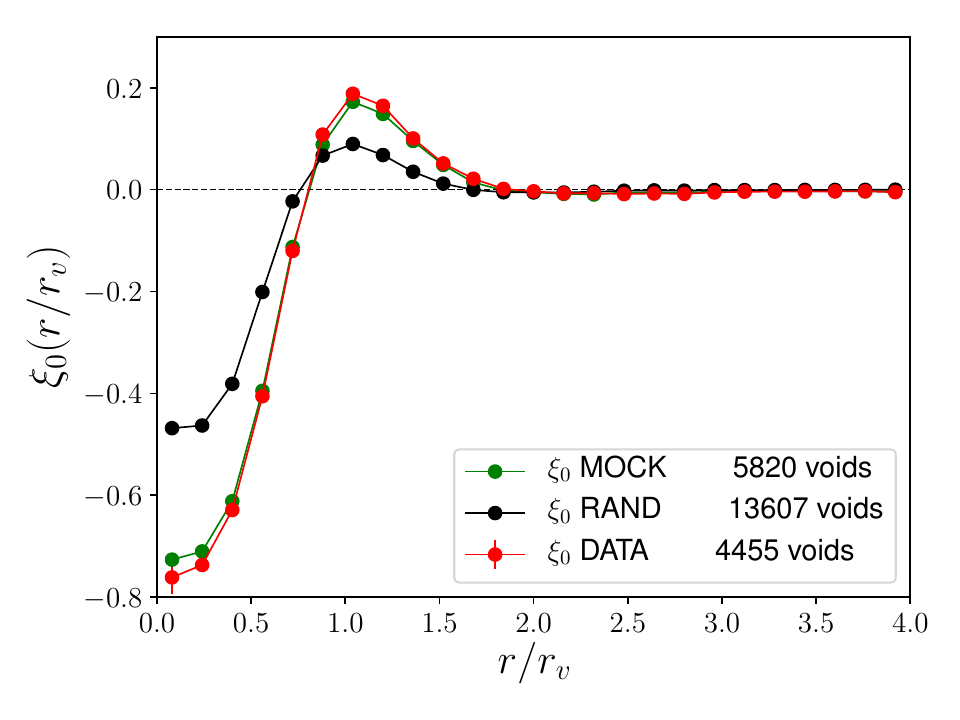}
	\includegraphics[width=\columnwidth]{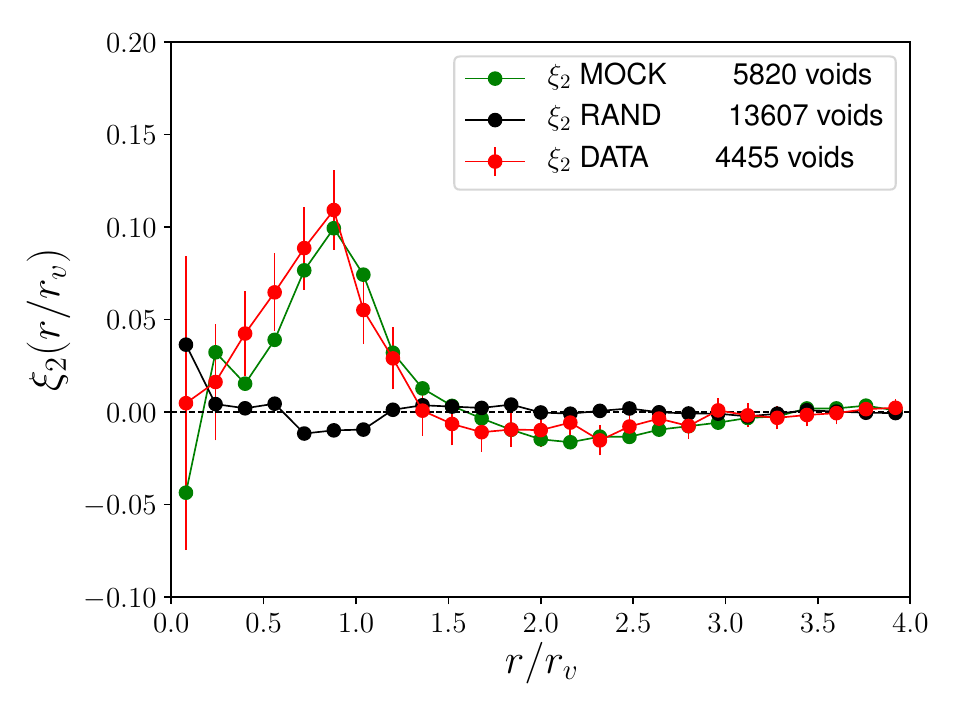}    
    \caption{Multipoles of the void-galaxy cross-correlation function. The left-hand panel shows the monopole $\xi_{0}$, the right-hand panel shows the quadrupole $\xi_{2}$, as a function of the void-tracer separation normalized by the void radius $r_{v}$. The green and red curves correspond to the mock voids and the data voids respectively. The black curves show the multipoles of the Poisson noise void-galaxy cross-correlation function. The errorbars are computed using the bootstrap method.}
    \label{fig:multipoles}
\end{figure*}

\begin{figure*}
	\includegraphics[width=\columnwidth]{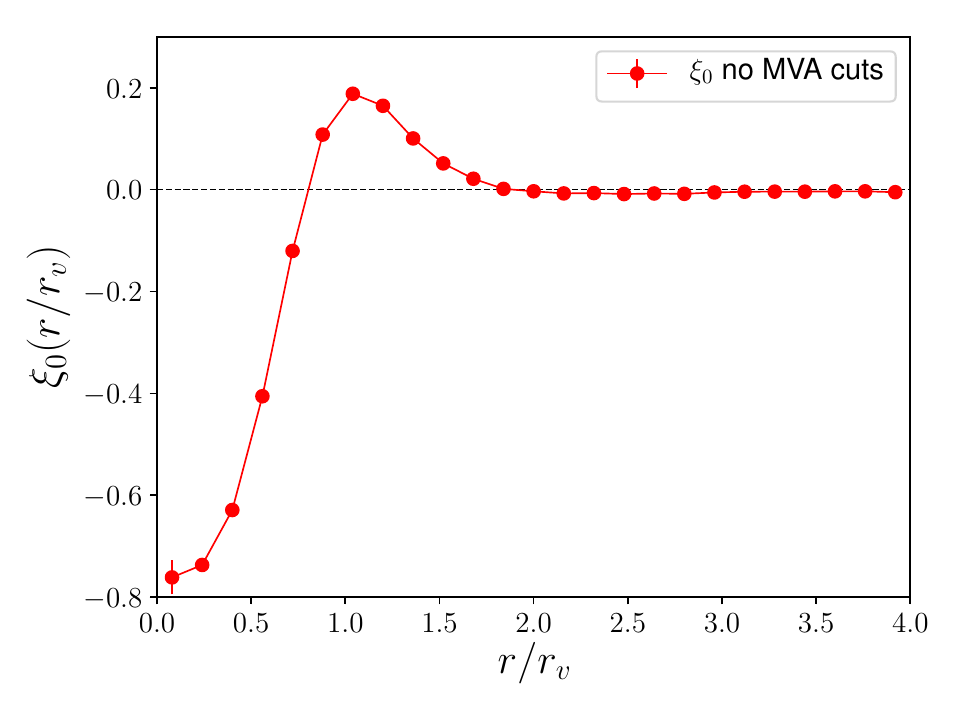}
	\includegraphics[width=\columnwidth]{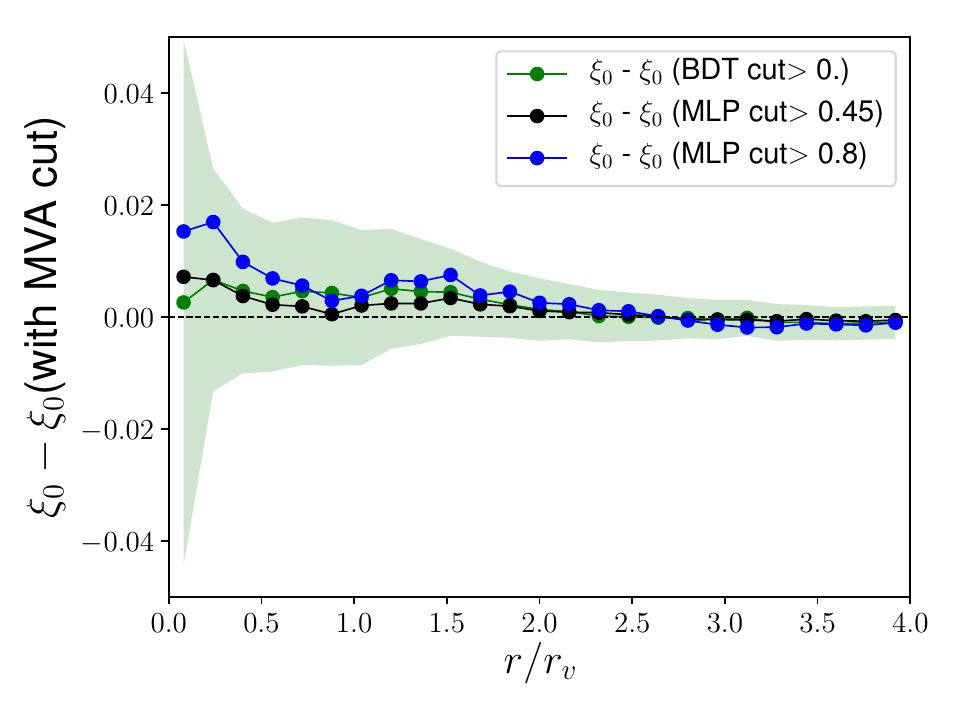}    
    \caption{Effect of the BDT and MLP response cuts on the monopole of the void-galaxy cross-correlation. Left-hand panel: the monopole $\xi_{0}$ as a function of the void-tracer distance normalized by the void radius $r_{v}$ for the full void sample ($4,455$ voids). Right-hand panel: differences between the monopole without any MVA cut and with a BDT or MLP cut on the response. BDT cut$\geq$0 in green ($4,060$ voids), MLP cut$\geq$0.45 in black ($4,142$ voids), MLP cut$\geq$0.8 in red ($3,603$ voids).  The green shaded area shows the effect of one sigma error on the difference between the monopole without cut and with the BDT cut.}
    \label{fig:monopoles_afterTMVA}
\end{figure*}

\begin{figure*}
	\includegraphics[width=\columnwidth]{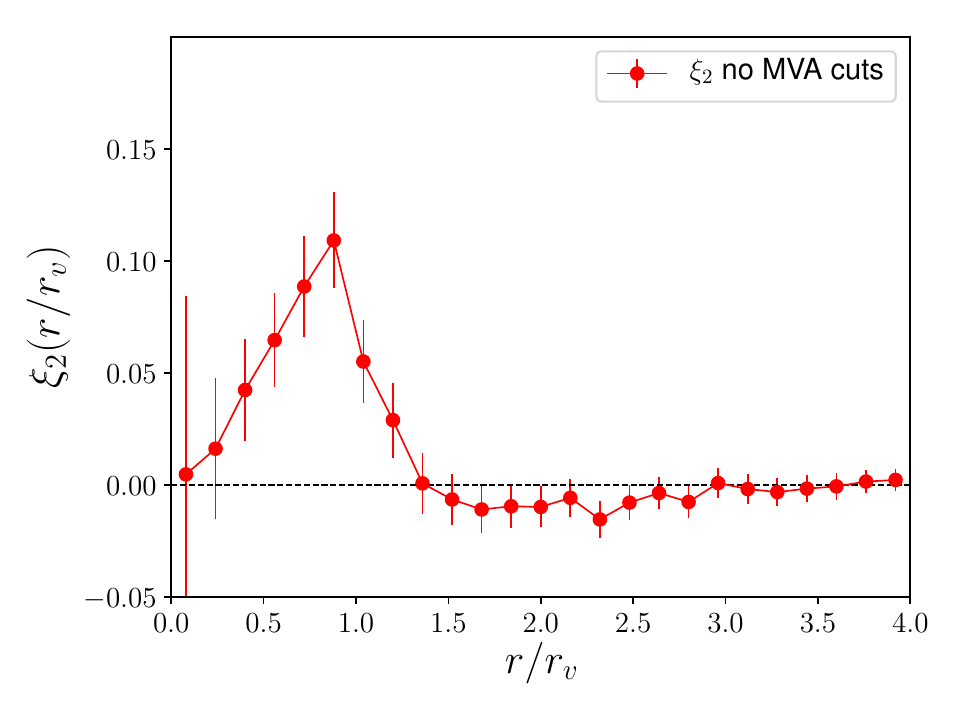}
	\includegraphics[width=\columnwidth]{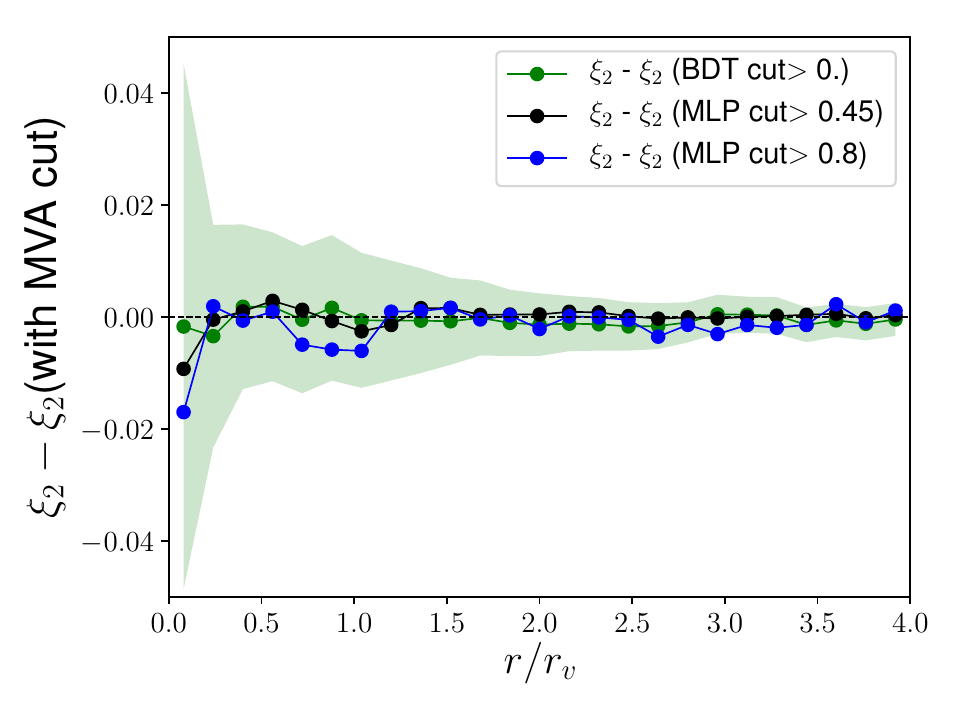}    
    \caption{Effect of the BDT and MLP response cuts on the quadrupole of the void-galaxy cross-correlation. Left-hand panel: the quadrupole $\xi_{2}$ as a function of the void-tracer distance normalized by the void radius $r_{v}$ for the full void sample ($4,455$ voids). Right-hand panel: differences between the quadrupole without any MVA cut and with a BDT or MLP cut on the response. BDT cut$\geq$0 in green ($4,060$ voids), MLP cut$\geq$0.45 in black ($4,142$ voids), MLP cut$\geq$0.8 in red ($3,603$ voids). The green shaded area shows the effect of one sigma error on the difference between the quadrupole without cut and with the BDT cut.}
    \label{fig:quadrupoles_afterTMVA}
\end{figure*}

\begin{figure*}
	\includegraphics[width=\columnwidth]{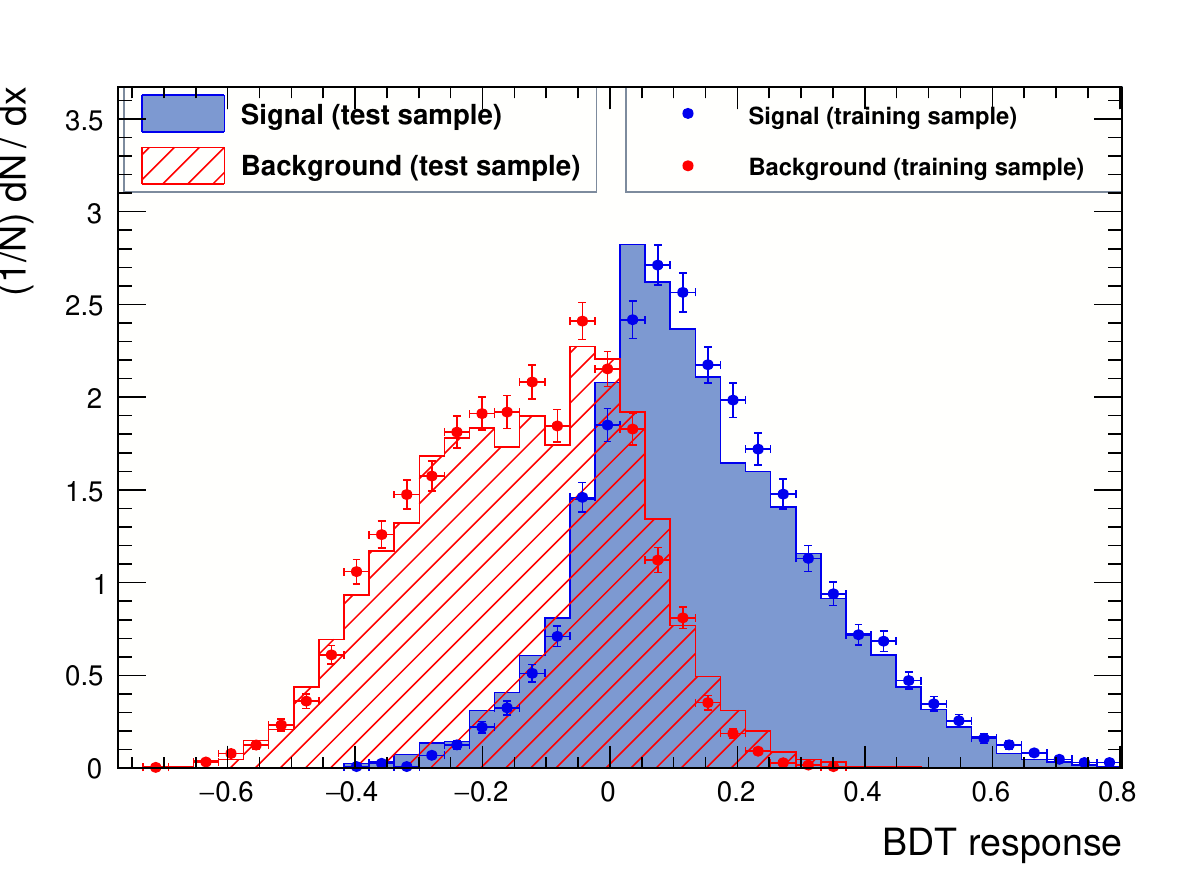}
 	\includegraphics[width=\columnwidth]{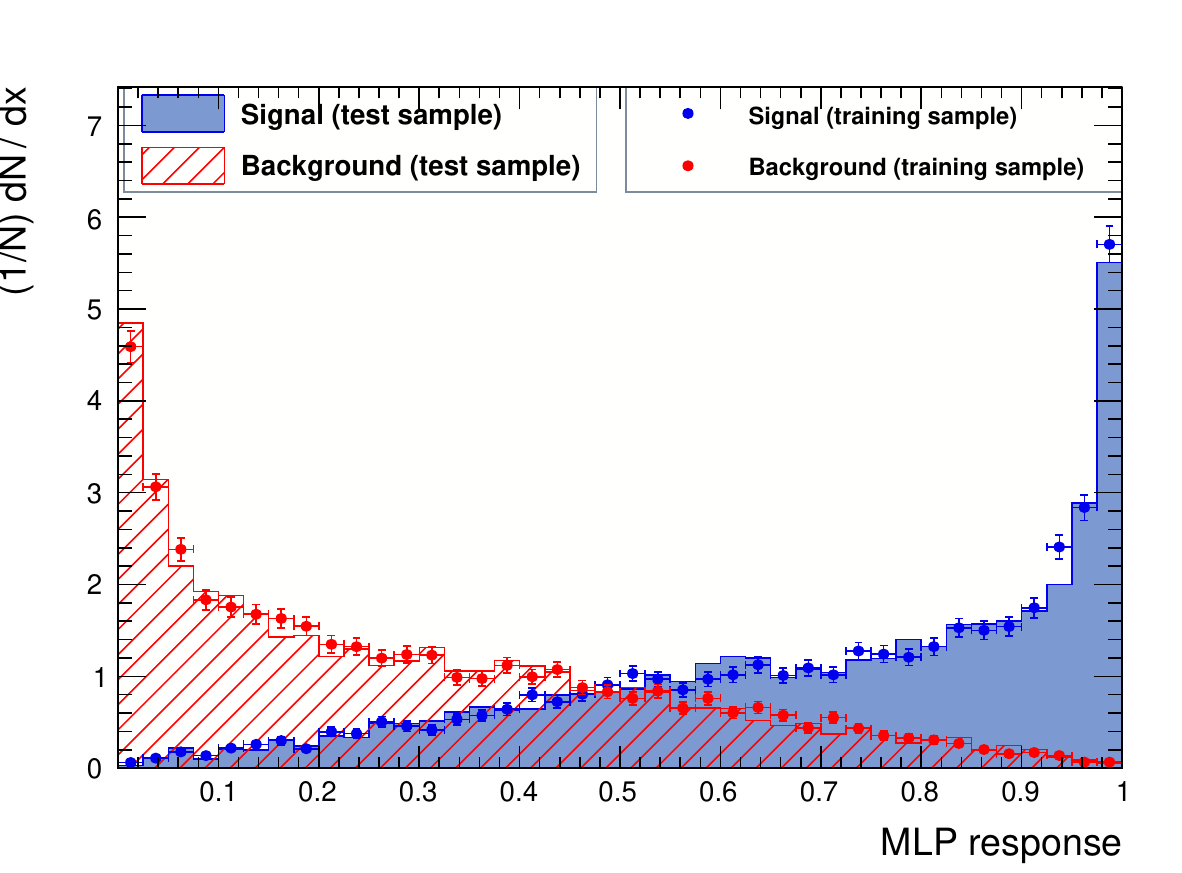} 
    \caption{Responses of the BDT and MLP classifications on the left-hand and right-hand plots respectively for the analysis of the sample of $100,000$ galaxies. The points with the error bars correspond to the training sample responses and the histograms to the test sample. The results for the signal are plotted in blue and those of the background in red.}
    \label{fig:100K_responses} 
\end{figure*}

\begin{figure*}
	\includegraphics[width=\columnwidth]{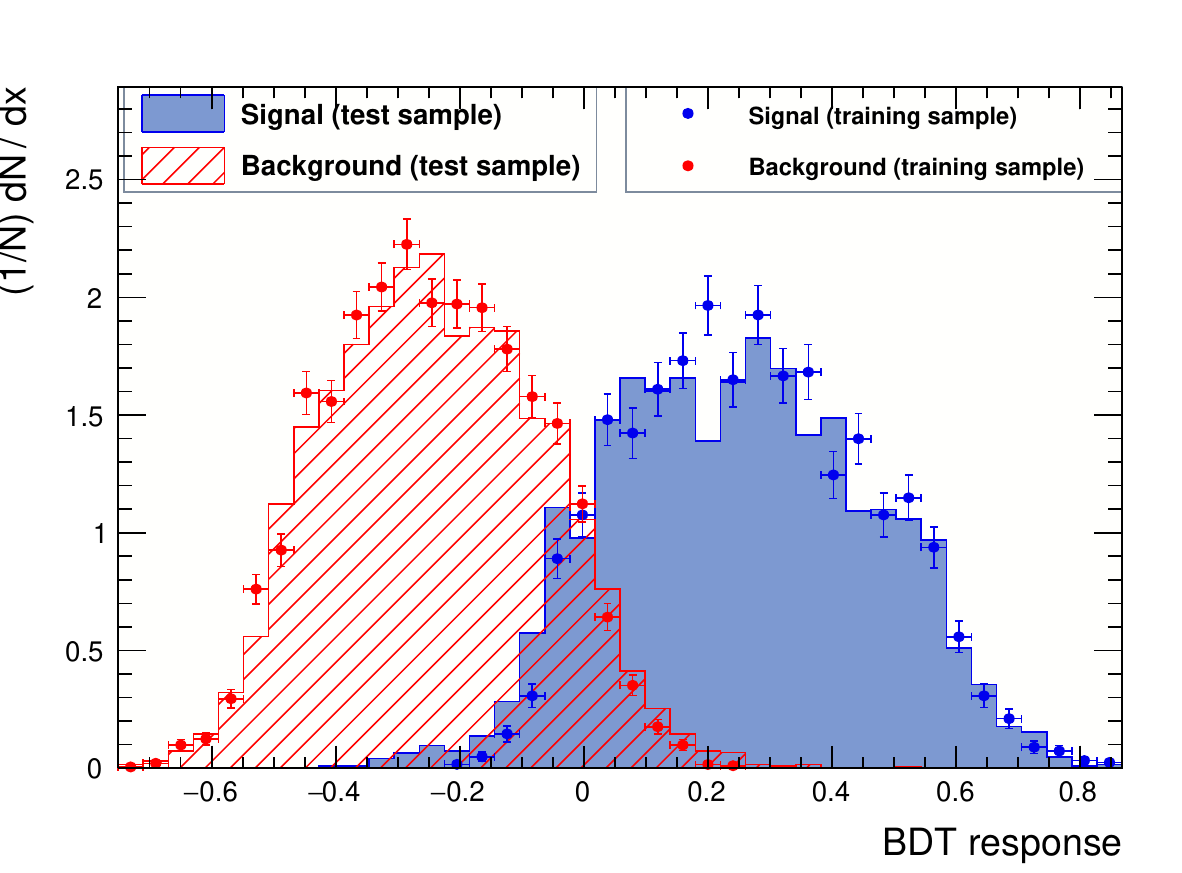}
 	\includegraphics[width=\columnwidth]{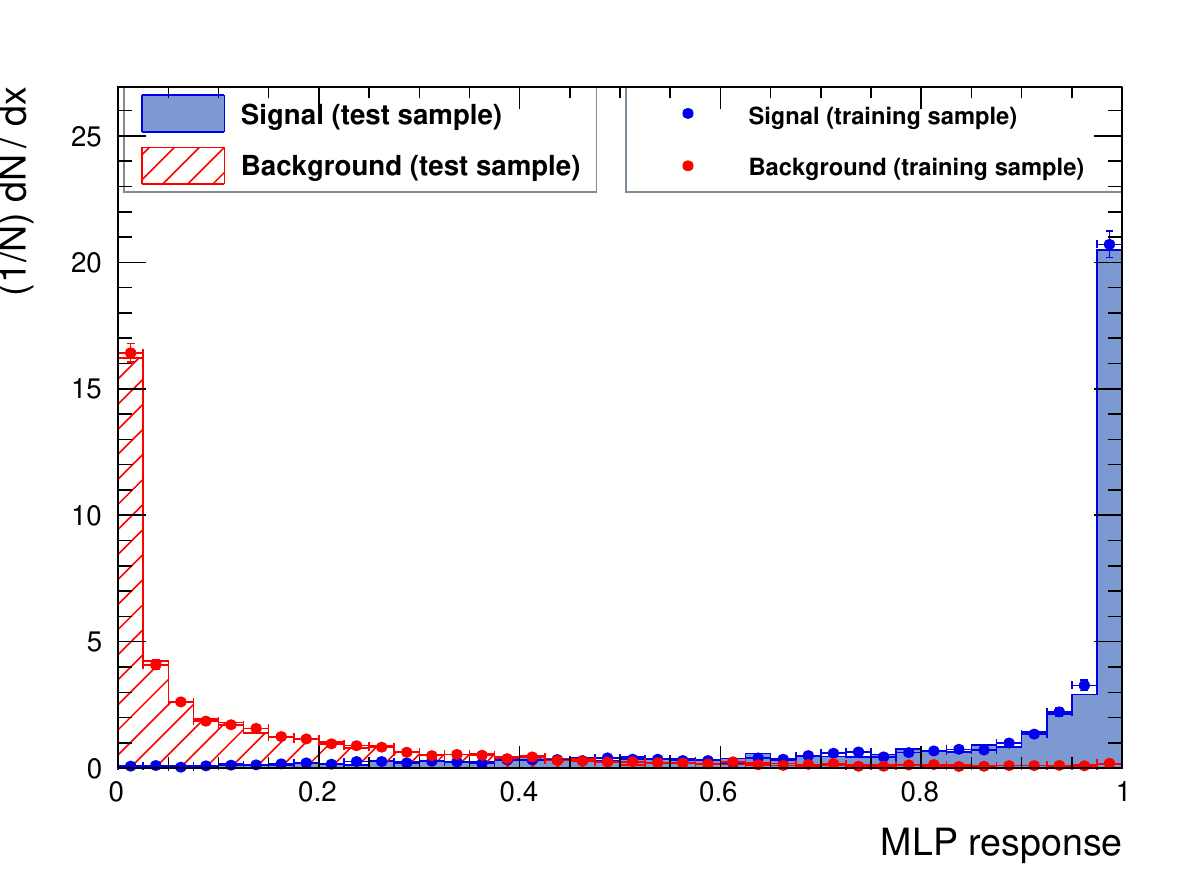} 
    \caption{Responses of the BDT and MLP classifications on the left-hand and right-hand plots respectively for the analysis of the sample with the cut on log($M^{*}$). The points with the error bars correspond to the training sample responses and the histograms to the test sample. The results for the signal are plotted in blue and those of the background in red.}
    \label{fig:MSTAR_responses} 
\end{figure*}

We compared these results with those obtained by applying simple cuts on the variables. 
In Fig.~\ref{fig:inputvar} we see that a combination of cuts such as: 
$\hat{V} \geq 10. ,  \rho_c  \leq1.5 ,  P(r) \leq 0.6 $ and $ N_t \geq 20.$ select signal type voids. Applying this set of cuts gives a signal efficiency equal to 53.5$\%$ with a background contamination of 4$\%$.  This shows the improvement brought by the MVA analyses which, with about the same background contamination, give a signal efficiency of 93$\%$ instead of 53.5$\%$.

The fact that this sample of voids is nearly free of Poisson type void contamination can  be interpreted in the following ways:

Firstly, this could be due to the watershed algorithm in VIDE. As discussed in section~\ref{S2} and shown in Fig.~\ref{fig:inputvar}, the Poisson noise voids in the SDSS BOSS DR12 CMASS galaxy sample would be identified as zones of higher core density and smaller density contrast. They would therefore be merged with the cosmic voids during the watershed transform step. 

Secondly, studies have been done to determine the best methods and conditions for the measurement of cosmological parameters with galaxy surveys~\citep{seo2003probing,feldman1993power,tegmark1997measuring}. In these papers it has been  shown that an important factor is the value of the product of the mean galaxy density $\bar{n}$ and the power spectrum $P(k)$. For shot noise to not significantly compromise the measurement,  the cosmic signal $P(k)$ should exceed the Poisson shot noise $1/\bar{n}$ and so the product $\bar{n} P(k)$ must be greater than $1$. In~\citet{seo2003probing}, the authors concluded that the value $\bar{n} P \approx 3$ is a good choice, being a good compromise between the reduction of the shot noise and the increase of the value of the error on the $P(k)$ measurement. ~\citet{reid2015sdss} show that in the SDSS BOSS DR12 CMASS galaxy catalog $\bar{n} P$ is approximatively equal to 3. Therefore the effect of shot noise is small. This could explain why there are very few voids from Poisson noise in the SDSS BOSS DR12 CMASS void catalog.

In section~\ref{sp}, we have shown that the efficiency of the MVA classification varies with the mean density of the tracer sample and with tracer bias. This comes from the fact that the differences between the distributions of the void characteristics in the signal and background samples are less discriminant. Since the SDSS CMASS sample contains almost no Poisson voids, we do not need to cut on the classifier responses to purify the sample, as shown in subsection~\ref{rsd}. However, analyses on voids extracted from sparser catalogs will benefit greatly from the MVA analyses. We will be able to enhance the number of reliable voids kept and reduce the number of spurious voids remaining in the signal by applying a cut on the classifier response. The BDT or the MLP responses could be converted to a probability, so another method to increase the fraction of genuine voids in a sample is to consider all the voids, but using these BDT or MLP probabilities as a weight.

\begin{figure*}
	\includegraphics[width=\columnwidth]{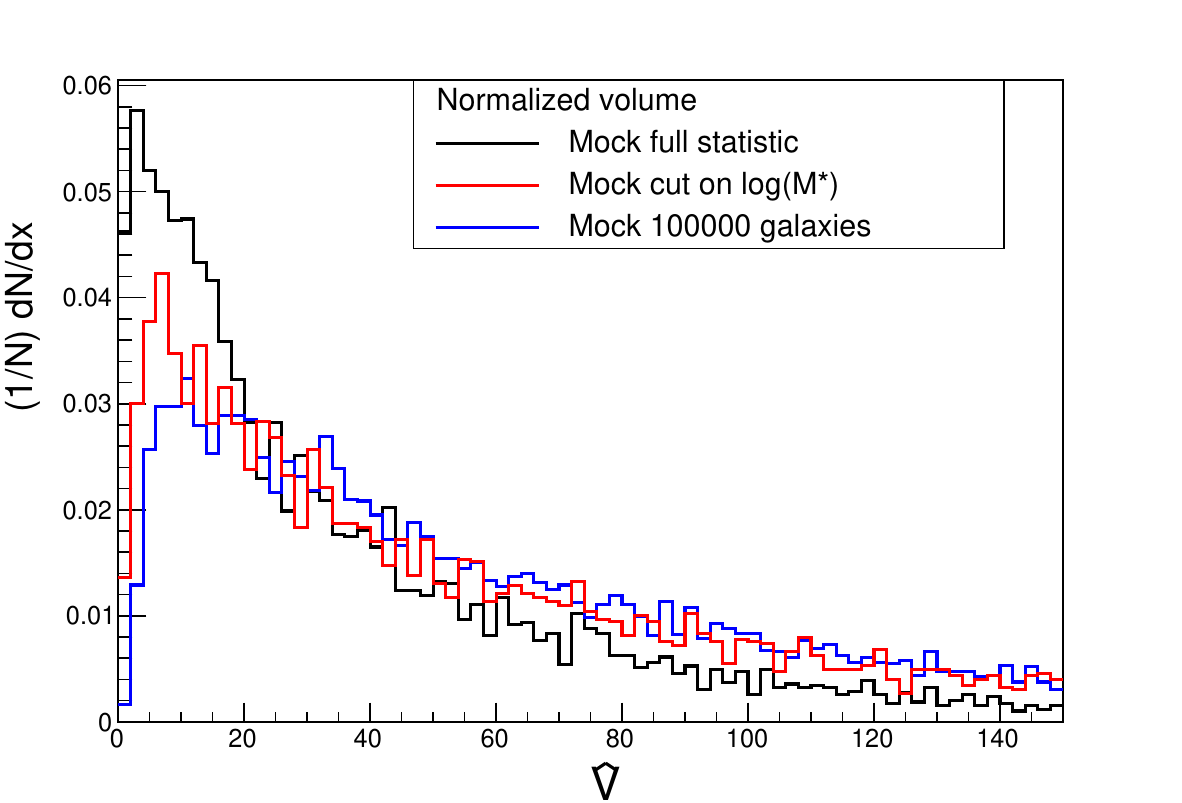}
 	\includegraphics[width=\columnwidth]{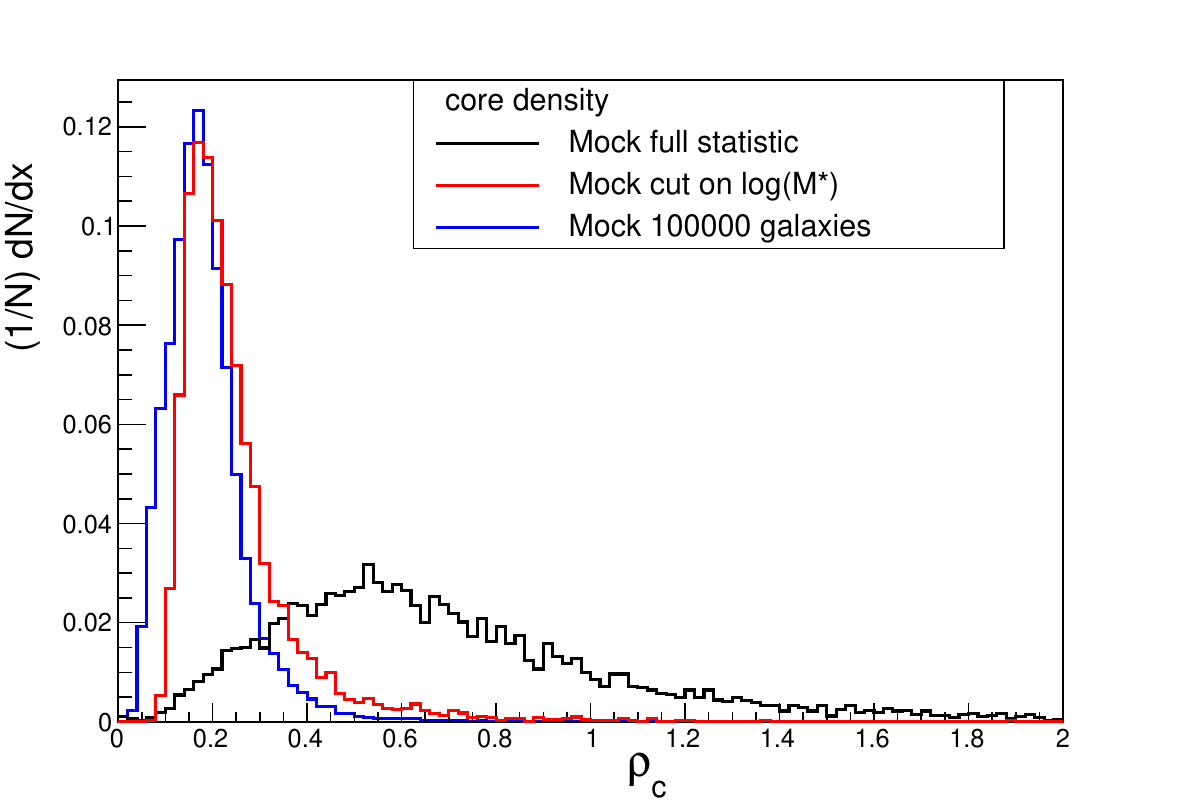} 
    \caption{Comparison of the distributions of the variables: normalized volume (left-hand plot) and core density (right-hand plot) of the full mock sample, the subsample of $100,000$ galaxies, and the sample obtained after the cut on log($M^{*}$)}.
    \label{fig:compvar} 
\end{figure*}

The mean and variance of the input variables could differ considerably from one sample to another. For example, the distributions of the normalized volume and the core density are shown in Fig.~\ref{fig:compvar} for the full mock, for the low density sample and for the sample with the cut on $M^{*}$. We note that the mean core density ranges from  0.67 for the full mock sample to 0.18 for the subsample of 100000 galaxies. The mean normalized volume value varies between 38.0 (in the full mock sample) to 60.5 (for the subsample of 100000 galaxies).  The weights during the training sample are optimized using cuts on the values of the variables. Therefore a specific MVA analysis should be performed for every void catalog. We conclude that, considering the high discriminating power of the MVA, it will be interesting to use this method to clean spurious voids from void catalogs obtained from upcoming cosmological surveys. Thus increasing the quality of cosmological constraints.
 
With the advent of upcoming large scale surveys such as DESI~\citep{Aghamousa2016}, Euclid~\citep{Laureijs2011} and WFIRST~\citep{Spergel2013}, larger volumes and higher galaxy number densities will be reached. This will allow us to construct very large void datasets. In the era of precision cosmology, aiming to produce robust void datasets is crucial in order to extract reliable cosmological measurements. The work presented in this paper ensures a robust validation of void catalogs for cosmological purposes and paves the way to increase the quality of constraints from cosmic voids in ongoing and upcoming cosmological analyses.

\begin{table*}
   \caption{signal efficiency and background contamination for the best cuts on BDT response or MLP response for the full sample}
	\begin{center}
          \begin{tabular}{ccccc}
            \hline
             Classifier & $\frac{S}{\sqrt{S+B}}$ & best cut value & signal efficiency $\%$ & bck. contamination $\%$\\
           \hline
            BDT & 30.1 & -0.02 & 95.8 & 5.3 \\
            MLP & 30.2 & 0.45 & 95.7 & 4.5 \\
           \hline
        \end{tabular}
       \end{center}
   \label{tab:effandbck}
\end{table*}

\begin{table*}
   \caption{Number of galaxies (sub-sampled from mock or randoms), mean number of voids per mock, mean number of voids per random sample. Signal efficiency and background contamination for the best cuts on BDT response or MLP response for each sub-sample of lower density}
	\begin{center}
         \begin{tabular}{cccccccc}
          \hline
             galaxies & voids (mock)  & voids (random) & classifier & $\frac{S}{\sqrt{S+B}}$ & cut value & signal efficiency $\%$ & bck. contamination $\%$ \\ 
             \hline    
              \multirow{2}{*} {200000}  & \multirow{2}{*} {1199} & \multirow{2}{*} {1961}  & BDT & 27.4 & -0.03 & 91.8 & 20.6 \\   
              &  & & MLP & 27.7 & 0.41 & 91.0 & 16.9 \\     
               \hline     
             \multirow{2}{*} {100000}  & \multirow{2}{*} {586}    & \multirow{2}{*} {815}   & BDT & 25.15 & -0.08 & 91.4 & 40.6 \\ 
              & & & MLP & 25.32 & 0.42 & 85.1 & 27.8 \\           
           \hline           
        \end{tabular}
   \label{tab:effandbck2}
  \end{center}
\end{table*}

\begin{table*}
   \caption{Number of galaxies (sub-sampled from mock or randoms), mean number of voids per mock, mean number of voids per random sample. Signal efficiency and background contamination for the best cuts on BDT response or MLP response for the sample with the cut on log($M^{*}$)}
	\begin{center}
          \begin{tabular}{cccccccc}
            \hline
             galaxies & voids (mock)  & voids (random) & classifier & $\frac{S}{\sqrt{S+B}}$ & cut value & signal efficiency $\%$ & bck. contamination $\%$\\
            \hline
            \multirow{2}{*} {194000}  & \multirow{2}{*} {1019} & \multirow{2}{*} {1911} & BDT & 28.4 & -0.04 & 92.9 & 14.0\\
            & & & MLP & 28.9 & 0.38 & 93.6 & 11.0 \\
           \hline
        \end{tabular}
       \end{center}
   \label{tab:effandbck3}
\end{table*}

\appendix
\section{}
When we observe objects like galaxies in  3-dimensional space, the radial distance to the object is determined by its measured redshift. The observed redshift comes from the Hubble flow, but also from a Doppler shift resulting from the peculiar velocity of galaxies.
This additional Doppler shift introduces a systematic distortion on the pattern of the distribution of galaxies in redshift space (as opposed to real space), an effect known as Redshift Space Distortion. In~\citet{hamaus2017multipole}, we measure the redshift-space distortion parameter $\beta$ (the relative growth rate) defined as $ \beta = f / b $,  where $f$ is the logarithmic growth rate for linear density perturbations and $b$ is the bias parameter.
This is done through the measurement of the multipoles of the 2-point correlation function (in our analysis, the cross-correlation function of voids and galaxies). 
If $\xi $ and $ \xi^{s} $ are the real-space  and redshift-space void-galaxy cross-correlation function respectively, we can write the decomposition of the redshift-space correlation function into multipoles using the Legendre polynomials $P_{l}(\mu)$: $ \xi_{l}(r) = \int_0^1 \xi^{s}(r,\mu) (1+2l) P_{l}(\mu) d\mu $, where $r$ is the distance between the void center and a galaxy and $\mu $ the cosine of the angle between the void-galaxy separation vector and the line-of-sight direction. The cumulative average monopole is defined as $ \bar{\xi_{0}}(r) = \frac{3}{r^{3}} \int_0^r \xi (r') r^{'2} dr' $.
Then $\beta$ can be calculated using the equation: $ \xi_{0}(r) - \bar{\xi_{0}}(r) = \xi_{2}(r) \frac{3 + \beta}{2 \beta} $ ~\citep{cai2016redshift}.

\section*{Acknowledgements}
M.C. C. thanks Yann Coadou (CPPM) for fruitful discussions on the MVA analyses, the TMVA tool and its utilization.

A. P. and A.J.H. acknowledge financial support of the OCEVU LABEX (Grant No. ANR-11-LABX-0060) and the A*MIDEX project (Grant No. ANR-11-IDEX- 0001-02) funded by the  Investissements d'Avenir french government program managed by the ANR. This work also acknowledges support from the ANR eBOSS project (under reference ANR-16-CE31-0021) of the French National Research Agency.

N. H. acknowledges support from the DFG cluster of excellence  `Origin and Structure of the Universe' and the Trans-Regional Collaborative Research Center TRR 33 `The Dark Universe' of the DFG.

Funding for SDSS-III has been provided by the Alfred P. Sloan Foundation, the Participating Institutions, the National Science Foundation, and the U.S. Department of Energy Office of Science. The SDSS-III web site is http://www.sdss3.org/.
SDSS-III is managed by the Astrophysical Research Consortium for the Participating Institutions of the SDSS-III Collaboration including the University of Arizona, the Brazilian Participation Group, Brookhaven National Laboratory, Carnegie Mellon University, University of Florida, the French Participation Group, the German Participation Group, Harvard University, the Instituto de Astrofisica de Canarias, the Michigan State/Notre Dame/JINA Participation Group, Johns Hopkins University, Lawrence Berkeley National Laboratory, Max Planck Institute for Astrophysics, Max Planck Institute for Extraterrestrial Physics, New Mexico State University, New York University, Ohio State University, Pennsylvania State University, University of Portsmouth, Princeton University, the Spanish Participation Group, University of Tokyo, University of Utah, Vanderbilt University, University of Virginia, University of Washington, and Yale University. 



\bibliographystyle{elsarticle-harv}

\bibliography{Void_characteristics} 

\end{document}